\documentstyle[12pt,aasms]{article}
\begin{document}
\title{ The Torque and X-Ray Flux Changes of \\ OAO 1657-415
 }
\author{ Altan Baykal }
\affil { Laboratory for High Energy Astrophysics
NASA/GSFC
 Greenbelt, Maryland 20771 USA
  NAS/NRC Resident Research Associate,\\
  Physics Department, Middle East Technical University,
Ankara 06531, Turkey }

\begin{abstract}

 Combining 
 previously published pulse frequencies and BATSE \\ mesaurements,
 we estimate 
the noise strengths (or power density estimates) 
of angular accelerations by using the root mean square 
residuals of angular velocity time series      
 of OAO 1657-415 and 
present the power spectra.  
The statistical interpretation of the angular velocity fluctuations 
are consistent with a random walk model. 
 In order to investigate the short term angular velocity   
fluctuations in detail, 
 a structure function analysis is applied for a two component 
neutron star model with a solid crust and a superfluid neutron core  
which is subjected to external white torque noise.
No evidence for core-crust coupling on timescales longer than one 
day is found. 
The correlations between X-ray flux and angular acceleration 
($\dot \Omega$)   
fluctuations are investigated. 
These are   
 compared with disk accretion 
theory (Ghosh $\& $ Lamb 1979 a,b) and
 wind accretion theory (Blondin et al., 1990).
 It is found that the most natural explanation
 of X ray flux and angular acceleration                         
 fluctuations is the formation of episodic accretion 
disks in the case of stellar wind accretion.

\end{abstract}
\newpage
\section{Introduction}

 The X ray source OAO 1657-415 was first detected by the Copernicus 
satellite (Polidan et al. 1978) in the 4-9 keV range. 
Initially association of this source with a massive star 
V861 Scorpii was considered. Subsequent observations by HEAO 1 
satellite (Bryne et al. 1979; Armstrong et al. 1980) and the Einstein 
Observatory (Parmar et al. 1980) did not confirm the 
association with V861 Scorpii. The HEAO 1 observations also 
showed 38.22 sec pulsations in the 1-40 keV and 
40-80 keV bands (White $\& $ Pravdo 1979; Byrne et al. 1981). Observations 
with Ginga and GRANAT 
 (Kamata et al. 1990; Gilfanov et al. 1991; 
Mereghetti et al. 1991; Sunyaev et al. 1991)  
 have shown episodic changes in the pulse period.
Timing observations of this source with the
 Burst and Transient Source Experiment (BATSE)  on the 
Compton Gamma Ray Observatory (CGRO) have 
shown that OAO 1657-415 is in an 11 days binary 
orbit with an X-ray eclipse by the stellar companion
 (Chakrabarty et al. 1993). The observed orbital parameters 
imply that the companion is a supergiant of 
spectral class B0-B6. 

 Massive X-ray binaries fall into three separate groups 
when the pulse periods are compared with orbital periods  
(Corbet 1986).  The systems with Be companions show  
correlations between the orbital and spin periods 
(Corbet 1986; Waters $\& $ van Kerkwijk 1989), while systems 
with OB supergiant companions fall into two separate 
regions. X-ray pulsars with Be companions 
 show transient behavior with  
episodic spin-ups (i.e EXO 2030+375, Parmar et al., 1989;  A0 535+26,  
  Finger et al., 1996)  
which suggest that the low velocity equatorial stellar wind 
forms an accretions disk during the periastron passages. 
 Systems with OB giants with pulse periods $P~<~10~$sec and orbital 
periods $P_{orbit}~<~4~$days  show 
optical photometric evidence for accretion disks (LMC X-1, Cen X-3, 
and SMC X-1) and nearly steady spin-up on longer time scales 
$>~10^{2}~ $days (Finger et al., 1993).
 BATSE observations of Cen X-3 have shown a lot of 
spin-up/down episodes on time scales $~<~10^{2}~$days, this behaviour in  
LMC X-1 and SMC X-1 is not known because of the long gaps between 
adjacent 
observations. Systems with longer spin periods $(>~100~$sec) 
showed short term spin-up/down episodes at $<~ 10~$days
 (i.e Vela X-1, see Deeter et al., 
1989) which can be explained in terms of
  flip flop instabilities  of wind accretion
(or very short time scale accretion disk formation; see  
Anzer et al., 1987; Matsuda et al., 1987;
 Blondin et al., 1990). OAO 1657-415 
falls between these groups in terms of orbital and spin period. 
In order to improve our understanding of the  
 spin history and the accretion 
process  
of the source,   
 in this work we study the statistical 
properties of the angular velocity changes 
and investigate the possible 
correlations of angular acceleration with X-ray flux.

 The angular velocity fluctuations in accretion powered 
 neutron stars are 
produced by torques originating outside and inside the 
object. The external torque is carried by the 
accretion flow; the internal torque depends on the coupling 
between the core superfluid and the solid outer crust. 
External fluctuations of the torque are filtered by the coupling 
between the crust and superfluid interior to produce an output 
represented by the observed changes in the angular velocity.
 The theoretical description of torque variations in terms of noise power 
spectral analysis was first studied by 
 Lamb et al., (1978a,b), who analyzed the response of a two 
component star to external fluctuations. Their model makes it possible 
to diagnose theoretically the properties of accretion 
flows and the internal structure of neutron stars. 
 Techniques for estimating the noise 
 power spectra in the case of nonuniformly  
sampled   
pulsar  timing 
 data were developed by Deeter $\& $ Boynton (1982) and  
Deeter (1984). They were applied to Vela X-1 and Her X-1 
  using the 
data obtained by HEAO-1 and UHURU  (Boynton 1981;
 Deeter 1981; Boynton et al., 1984; Deeter et al., 1989). 
The results showed that the angular velocity time series 
 of Vela X-1 and Her X-1 can be 
modelled as a random walk (or white noise in the 
angular accelerations with a power law index 
$n=0$). Recent results of BATSE observations 
 are indicating that 4U 1626-67 has an angular velocity time series 
 which is also consistent with the random walk model 
 (Chakrabarty et al., 1995). 
 Random walk in angular velocity   
can be characterized by a rate of torque events $R$ of 
steps $\delta \Omega $. The rate and the step size of the 
events depend on the type of accretion (Prince et al., 1994). 
For example in the wind accretor Vela X-1,
 the spin-up/down time 
scales are around day (Deeter et al., 1989), therefore  
the rate of the torque events 
may be $R < 1~$days$^{-1}$ while in Her X-1 
the rate is lower due to the 
disk accretion (Wilson et al., 1994).  
   Time domain techniques such as 
autoregressive time series (Scargle 1981) were applied 
by Baykal $\& $ \"{O}gelman (1993) and de Kool $\&$ Anzer (1993) in order 
to model the angular velocity time series of accretion powered X-ray binaries. 
They found that some of these systems have 
angular velocity time series  
which are consistent with random walk. 
Recent BATSE results 
for Cen X-3 and GX 1+4 
 have shown that the noise power spectrum is redder then the random walk 
model in angular velocity time series   
(or flicker noise in the angular 
accelerations with power law index    
 $n=-1$; see 
Finger et al., 1994; Chakrabarty et al., 1995).
The reason for flicker noise in the angular accelerations   
may be the smooth  transitions from  
spin-up to spin-down (or vice versa) 
(see Finger et al., 1994)
 which makes the noise power spectra steeper (or redder).

The sign and magnitude of the torque physically depends on the 
magnetosphere of the neutron star
 and the type of accretion flow exterior to 
magnetosphere (Ghosh $\& $ Lamb 1979 a,b; Blondin et al., 1990).
 One way to study the
accretion dynamics of neutron stars is to observe the nature
of the correlation between changes in rotation rate
($\dot \Omega $) and mass accretion rate ($\dot M$)
(or the related quantity 
 X ray flux ). 
Such a study may provide more detailed information 
on the physics of accretion and helps to discriminate
between disk and wind-type accretion 
 (Ghosh $\& $ Lamb 1979 a,b; 
Blondin et al., 1990). 

 In this paper, we apply various techniques in order  
to study the torque fluctuations  
 of OAO 1657-415. In Section 2, the data base used in this 
analysis is described.   
In Section 3, we construct 
the low resolution 
 power density spectrum using the mean-squared residuals 
technique developed by Cordes (1980) and Deeter (1984). 
 In Section 4, to see the sharp changes of 
rotation rate, a structure function is calculated 
Cordes (1980); the two component neutron 
star model and the response of the neutron star to external  
white torque noise is compared with the observed data. 
In Section 5, the correlations between 
the angular acceleration ($\dot \Omega $), 
 X-ray flux, and specific angular 
momentum  
 are studied. 
These correlations are compared with the disk accretion hypothesis 
(Ghosh $\&$ Lamb 1979 a,b) and the wind accretion hypothesis 
(Blondin et al., 1990).

\section{The Data Base} 

 In this work, we used pulse frequency records   
 compiled from the published literature (Nagase 1989; Gilfanov et al. 1991) 
and generated from BATSE observations.  
 The total time span 
of the observations covers approximately $16~$yr.  
 The measured angular velocities are plotted as a function of time in Fig. 1. 
The BATSE detections of angular velocities are plotted in  
 Fig. 2 . These records are measured on the basis of daily analysis   
after removing binary motion of the neutron star and  
 are available through anonymous ftp from HEASARC.   
The BATSE data base contains 1219 pulse frequency 
 and pulse flux measurements of OAO 1657-415 
 in a time span of over 3 years. 
(BATSE instrument has 8 uncollimated detector
modules arranged on the corners of the CGRO spacecraft
 which are sensitive $>20$keV; for details of the observatory and
 observation
 see Fishman et al., 1989; Chakrabarty et al., 1993).
The gaps in Fig. 2 corresponds to either 
no detection of the source or unsignificant 
detections.

\section{ Power Spectra     }

 The root mean square residuals 
  technique used in this section for the estimation of red 
noise power density and associated random walk noise strengths 
is discussed in detail by Cordes (1980) and  Deeter (1984). 
The applications made by Baykal et al., (1993a,b) are the examples 
of its use in characterizing random walk processes.  
 
  For the case of $r=1,2..$th-order red noise 
 (or $r$th-time integral of white noise time series)
 with strength $S_{r}$, the 
mean squared residual for data spanning an interval $T$ 
is proportional to $S_{r}T^{2r-1}$. The proportionality 
factor (or normalization coefficient)
 depends on the degree $m$ of the polynomial removed 
prior to computing the mean square residual; this factor can be obtained 
by determining the expected mean square residual for unit strength red noise 
($S_{r}=1$) over a unit interval ($T=1$), either by Monte-Carlo 
methods (Cordes 1980) or by mathematical evaluation (Deeter 1984). 
The mean square residual, after removing a polynomial of degree 
$m$ over an interval of length $T$, is then given by
\begin{equation}
<\Delta \Omega ^{2}(m,T)>=S_{r}T^{2r-1}<\Delta \Omega ^{2}(m,1)>
\end{equation}
where $<\Delta \Omega ^{2}(m,1)>$ is the proportionality 
factor which has been 
derived from a unit-strength noise process. 
Our preliminary analysis showed that the angular velocity residuals 
can be characterized by first order red noise (or random walk).   
This noise process is adopted in computing the normalization coefficients. 
The normalization coefficient is computed by Monte Carlo simulation,
sampling the simulated data as the real data set was sampled (Cordes 1980).
Because most
of the BATSE data are approximately equally spaced at 1 day intervals,
the normalization coefficient differs by no more than 15$\%$ 
from the theoretical
prediction for equally spaced data (assuming a sampling interval of one day)
as given in Table 1 and 2 of Deeter 1984.
 
As a next step, time scales are sampled 
 at nearly octave spacings for the power density estimators (Deeter 1984).
To do this the entire length of the data is taken as the longest 
time-scale. Then the intervals are halved successively to get down 
to the shortest practical time scale. Intervals with insufficent 
data points are discarded. The mean squared residuals for the
 remaining 
intervals are calculated and converted into the noise strengths
($S_{r=1}$) using Eq. 1. 
The contributions of the measurement errors have been  
computed   
 by converting the estimated variances of angular velocities  
 into noise strengths and by employing the identical procedure above.

Estimated noise strengths on each time scale are combined into a 
single power density estimate
( $P_{\dot \Omega} = (2 \pi f)^{2} P_{\Omega }=S_{r=1}$, where 
$f$ is the frequency, Rice (1954);  Boynton (1981))  
 by averaging, and are represented by    
a frequency equal to the inverse of the time scale. 
In order to estimate the power density error bars, 
the effective number of degrees of freedom, for each estimate 
is determined by generating its statistical distribution by using 
Monte Carlo simulations of a first order red noise process, sampled 
in the same way as our angular velocity data 
(Blackman $\&$ Tukey 1958; Deeter $\&$ Boynton 1982).
  Then the power density error bars are  
computed 
for a given effective number of degrees of freedom by computing 
the power density values corresponding to the $16\%$ and $84\%$ points 
of the $\chi ^{2}$ distribution; these are equivalent to $\pm 1\sigma $ 
confidence level for a Gaussian
(Blackman $\&$ Tukey 1958). In Fig. 3, we display the resulting 
power density estimates (or noise strengths) of the angular acceleration  
 fluctuations.  

 In the analysis, we have found the crossover frequency between the 
 errors and 
 noise strengths around $ f_{c}~\approx 1/(8~$days).  
 We have fitted a straight line 
 to the 8 lower octave power density estimates and found  
 $n = -0.02 \pm 0.3 $ which is consistent 
 with white noise $n \approx 0$ in the angular accelerations ($\dot \Omega $)
 with the mean power density 
(or noise strength)
 $S=R<\delta \Omega ^{2} >=8.1 \pm 1.2 \times 10^{-16}$ rad$^{2}$/sec$^{3}$.

 In order to crudely estimate the step size of the fluctuations 
 $<\delta \Omega ^{2} >^{1/2} $ and the rate $R$   
 of torque  events we make use of the autocorrelation of 
angular acceleration $(\dot \Omega $) time series.  
 We took the numerical time derivative of the  
angular velocity ($\Omega $) time series 
at time scales corresponding to the cross over frequency (see Fig. 7b)  
  and represented  
 the autocorrelation function in Fig. 4. 
 Autocorrelation function of white noise time series 
 can be expressed as (Jenkins $\&$ Watts 1968),
\begin{equation}
<\delta \dot \Omega (t) \delta \dot \Omega (t')> =
 \sigma _{W} ^{2}  \delta (t-t')
\end{equation}   
where $\sigma _{W}^{2} $ is the white noise variance 
 at zero time lag $t=t'$. As seen from Fig. 4, actually 
 the autcorrelation function shows correlation for time 
 lags ($\Delta t $) up to 8-19 days.
    
 The step size of the 
 fluctuation during this time scale can be estimated \\
 $<\delta \Omega ^{2}> \approx  <\delta \dot \Omega >^{2} 
 \Delta t ^{2}$. By using the values
 $<\delta \dot \Omega >^{2} \sim 7.7 \times 10^{-22}$ rad$^{2}$/sec$^{4} $ 
and $\Delta t $, from the 
autocorrelation function and  
the random walk noise strength estimate 
from the power spectrum we estimated the rate of torque 
events as
\begin{equation}
R=\frac{S}{<   \delta \Omega ^{2} >}  \sim 
1/30 - 1/5~ {\rm days^{-1}}. 
\end{equation}   

 This implies a torque event occurs within  
 $1/R \sim 5-30$  days and lasts    
  $\Delta t \sim  8-19 $ days.
 These events appear as a delta function 
(or unresolved events) if we are studying the
long time scale  fluctuations
 and make the power spectra
 of $\dot \Omega $ 
 flat at low frequency (or long time scales $t >> \Delta t~$days ).
 But at high frequencies (or short time scales $t \sim \Delta t~$days), 
 it is quite natural to see decrease in power density estimates. 
 If we express the white noise time series as a sum of 
 Gaussians      
$
\sum_{i} \delta \Omega _{i} \frac{1}{(2 \pi )^{1/2} \sigma }
 e^{-(t-t_{i})^{2}/(2\sigma ^{2})} $,
 with a small finite  
  variance 
 $\sigma ^{2}  \sim \Delta t ^{2}$, then we would 
 expect the power spectrum to have the form \\ 
$P_{\dot \Omega }  \sim R<\delta \Omega ^{2}>
 e^{-2f^{2} \Delta t ^{2}}$. Therefore at frequencies   
 $f \sim 1/ (2)^{1/2}\Delta t  $ one should  
  see a decrease 
 in the power spectra at order of 
  $\sim e^{-1}$.  
  This is qualitatively seen 
 in our low resolution power spectra (see Fig. 3).

\section{Structure Function Analysis }

 In this section we study the short term fluctuations of 
 angular velocites by employing structure function analysis 
 (Lindsey $\& $ Chie 1976; Cordes $\& $ Downs 1985). 
 This technique is more sensitive to sharp fluctuations.  Therefore 
 it is more likely to show the effects of a possible decoupling between the 
crust and the superfluid core.  
 A first order 
 structure function for the angular velocity time series 
 can be defined as 

\begin{equation}
 D_{\Omega}  (t_{o},t) =<(\Omega (t_{o}+t)-\Omega (t))^{2} >.
\end{equation}

 For a random walk in frequency one can express the time series 
 as  
\begin{equation}
 \Delta \Omega (t) = \sum_{i} \delta \Omega _{i}  \theta  (t-t_{i} )
\end{equation}
where $ \delta \Omega _{i}$ is the step size of events
 $\theta  $ is the step function and $t_{i}$ is the events occur at 
random times with a rate $R$. The structure function of the 
above equation can be written as (Cordes 1985), 
\begin{equation}   
 D_{\Omega}(t)=<\Delta \Omega ^{2}>=R<\delta \Omega ^{2} > t = S t  
\end{equation}
 where $t$ is the lag between two measurments of $\Omega $, 
$S$ is the random walk noise strength. 

 If the neutron star has significiant superfluid in the core 
 (Lamb et al., 1978a,b; Sauls 1988; Lamb 1991; Datta $\&$ Alpar 1993)   
 then 
 the moment of inertia of a neutron star resides mainly in its neutron 
 superfluid core. There is a neutron superfluid also in the 
 inner part of the crust lattice. This crust neutron superfluid 
 carries $10^{-2}$ of the stars's moment of inertia, and is coupled 
 to the rest of the crust on very long time scales, typically 
 extending to years (Alpar et al., 1981; Alpar et al., 1993). 
 The crust superfluid moment of inertia is resolved  
 in radio pulsars which are spinning down due to electromagnetic 
 dipole radiation,     
 as a sudden changes in rotation rate (or glitch events) with 
a magnitude 
 $\Delta \Omega / \Omega \sim 10^{-9}-10^{-6} $
 and its time derivative
$ \Delta \dot \Omega / \dot \Omega \sim 10^{-3}-10^{-2} $   
   (Lyne 1993; Shemar 1993). 
 These kind of
  glitches can not be detected in the accretion powered X-ray binaries 
 because the external torque noise in these systems  
 dominates 
  (Baykal $\&$ Swank 1996). Also the internal dynamics of  
  radio pulsars spinning down may be quite different 
 (Alpar 1993). In this work we will not be 
 concerned with the crust superfluid. In the core of the neutron star,
 the rotating neutron 
superfluid is coupled to the stellar crust by interactions 
between the charged particles (electrons and protons) and the 
quantized vortices. In the case of 
 any change in rotation velocity of the crust the whole charged 
 component (protons, electrons) of the stellar interior is coupled 
 to the crust on a short time scale, typically less than 
 $10~sec$ (Easson 1979). The angular velocity 
of the neutron superfluid is determined by the density 
of quantized vortices. The rotation velocity 
of the core neutron superfluid will not change unless the 
quantized vortex lines move radially (outward for spin-down and inward 
for spin-up). Any fluctuation on the crust creates a force 
on the vortices because of the relative velocity between the 
vortex line and charged component. Vortex lines relax with charged 
particles at dynamical 
(or at the crust core coupling time $\tau $) time scales, which is of the order of 
$\tau = 100 (m/\delta m)^{2}~P \sim 10^{2}-10^{4}~P $,
 where $P$ is the rotation period of the neutron star, $\delta m$ is the 
change in mass of proton as a result of its coupling with 
neutrons  
(Alpar et al., 1984; Alpar $\&$ Sauls 1988).   

 The above senario can be approximated with a two component 
 neutron star model (Baym et al., 1969). 
 In this model, one component is the 
 crust-charge particle system, which consists of protons, 
 electrons and the crust with an inertia $I_{c}$ with rotates 
with angular velocity $\Omega _{c}$. 
 The second component is the core neutron superfluid, with 
 moment of inertia $I_{s}$, with rotates with angular 
 velocity $\Omega _{s}$.
 Any external torque on the crust creates a lag between $\Omega _{s} $ 
 and $\Omega _{c} $.  
The two components are coupled
 by a crust core coupling time $\tau $, 

\begin{equation}
 I_{c} \dot \Omega _{c} = N(t) - \frac{I_{c}}{\tau }
 (\Omega _{c} - \Omega _{s}) ,
\end{equation} 

\begin{equation} 
I_{s} \dot \Omega _{s} = \frac{I_{c}}{\tau } (\Omega _{c} -\Omega {_s}).
\end{equation}
Here $N(t)$ is the external torque exerted on the star. From the 
power spectra in the 
previous section we found that  
  $\dot \Omega $ fluctuations are consistent with 
white torque
 noise. We assume that external torque fluctuations are also white noise
$N(t) = \sum \delta L_{i} \delta (t-t_{i})$, where
 $\delta L_{i}= I \delta \Omega_{i} $ 
is the angular momentum added or subtracted from the star at $t=t_{i}$. 
Then the fluctuations of the crust can be expressed 
 (Baykal et al., 1991) as  
\begin{equation}
\Delta \Omega _{c} =
\sum _{i} \delta \Omega _{i} \theta (t-t_{i}) 
(1 + \frac{ I_{s}}{I_{c}} e^{ (-(1/\tau)(I/I_{s})(t-t_{i})} ),
\end{equation}
where $I=I_{s}+I_{c}$ is the moment of inertia of the neutron star.
If the neutron star response purely rigid $t >> \tau $ or 
$I_{s} \sim 0 $ then the time series behave as a pure random walk 
in angular velocity     
(see Eq. 5). 

By using the definition of 
 the structure function for the time series above and 
 defining the $\Delta \dot \Omega = \Delta \Omega _{c} / t$,
 then the mean square fluctuation of the white noise 
 variable can be written as 
\begin{equation}
<\Delta \dot \Omega ^{2}>=\frac{R<\delta \Omega ^{2}>} 
{t} (1+\frac{\tau '}{t}\frac{I_{s}}{I_{c}}(1-e^{(-t/\tau ')}))^{2}
\end{equation} 
where 
  $\tau ' = \tau I_{s} /I $.  At long time lag $t >> \tau '$,  
 angular accelerations fluctuations  
 behaves as   
$ <\Delta \dot \Omega ^{2}>^{1/2}=(R<\delta \Omega ^{2}>/t)^{1/2}$, 
while at short time lag  $t << \tau '$ behaves as 
$<\Delta \dot \Omega ^{2}>^{1/2}=
I/I_{c} (R<\delta \Omega ^{2}>/t)^{1/2}$. 
This means that in the long time lag limit, the core superfluid 
couples to the neutron star crust and responds to the external torque 
with total moment of intertia $I$. On the other hand in the short time 
lag neutron star responds to external 
fluctuations with crust moment of inertia 
 $I_{c}$ (or with charged particle components) 
 (see also Lamb et al., 1978a,b). 
 In Fig. 5a, we plotted the angular acceleration 
fluctuations in the case of core superfluid mixture 
 $I_{s}/I_{c}= 0, 3, 10$ for a crust 
core coupling time $\tau = 1~$day. In the plot we adapted the 
noise strength value ($S$) from the power spectrum 
as $S =R<\delta \Omega ^{2} >= 8.13~ 10^{-16}$ rad$^{2}$/sec$^{3}$.  
As it is seen from Fig. 5a, if there is a significant 
core superfluidity and even if the      
  crust core coupling time is 
relatively short $\sim 1~$day, 
 the crust response increases the 
 angular acceleration fluctuations  
 up to time lags  
 of tens of days . In other words even if we do not see the torque  
events on shorter time scales 
 we can resolve the crustal moment of inertia 
 by measuring the angular accelerations and testing with a 
simple two component 
 neutron star model.  
 In Fig. 5b,c, we simulated 1000 independent time series in the 
form of Eq. 9,  
 is 
sampled  according to the  
 OAO 1657-415 angular velocity history (see Fig. 2). Then we compared the 
angular acceleration fluctuations with observed fluctuations. 
In the simulation,  
we sampled the events uniformly with a rate $R \sim 1/18~$days$^{-1}$ and used 
the input variance
 $<\delta \Omega ^{2}> = S/R \sim 1.25~10^{-9}$~rad$^{2}$/sec$^{2}$.
 First, we simulated the time series   
with $I_{s}/I_{c}=10,$ and $ \tau =1~$day;  
 the simulated angular frequency 
fluctuations (with larger error bars)   
are shown together with the observed fluctuations (with smaller error bars)  
in Fig. 5b.
 Clearly  
the simulated fluctuations are not compatible with the observed 
fluctuations. In Fig. 5c, a pure 
random walk time series (or  
a rigid body respose with  $I_{s}/I_{c}=0$) is simulated. In this case,   
the simulated and observed angular frequency  
  fluctuations agreed with each other at $\sim 1 \sigma $ level 
at all time lags  (see Fig. 5c). 
This indicates that
 the angular acceleration fluctuations seen in OAO 1657-415,  are
associated with external torques.
The absence of a signature 
 of core superfluidity is suggesting us that either crust-core 
coupling time is so short (i.e. $ \tau~<<~1~$days $\sim  2273 P  $ 
 where $P=38~$sec is the pulse period) that   
 all charged components and the core superfluid 
  couple on the order of hours or the core superfluidity 
is not significant $I_{s} <<  I_{c} $.
 If the latter  
possibility is correct, 
 this is suggesting that either      
 most of the neutrons in the core are too hot 
to be in the    
 superfluid phase  (Ainsworth et al., 1989) 
or the equation of state is stiff with higher crustal 
moment of inertia (Lamb 1991).   
The well studied high mass X ray binary Vela X-1 showed that
less than  
$ 85\%$ of the moment of inertia of the star is weakly 
coupled to crust with coupling times in the range 
$\tau \sim (305 - 9159) P $, 
 P=283 sec  (Boynton et al., 1984; Baykal et al., 1991) 
while radio observations constrain  
the crust core coupling time to be $\tau ~<~1350P $ (Chau 1993).

\section{The correlations between Angular Acceleration,
 Flux and Specific Angular 
Momentum }
 In this section, we study the various physical 
 correlations between angular acceleration and mass accretion rate 
 (or pulse flux). 
 The BATSE instrument has a response 20-60 keV (Fishman et al., 1989). 
 Therefore the pulse flux 
 time series which is represented in Fig. 7a,   
 may not represent the bolometric X-ray flux 
 (White et al., 1983). 
 The bolometric X-ray flux $L_{x}~ erg/sec$ is related to mass accretion 
 rate $\dot M$,  
\begin{equation} 
L_{x}= \eta ~G~M \dot M /R
\end{equation}
where $\eta \approx 0.1$ is the efficiency factor,
 $G$ is the gravitational constant, $M$ is the mass of neutron star,  
$R$ is the radius of the neutron star.     
In order to understand qualitatively whether the X-ray pulse flux 
fluctuations are related with mass accretion rates,  
we represent the autocorrelation function of pulse flux time series 
in Fig. 6.  As it is seen from the Fig. 6,  
the pulse flux fluctuations are becoming uncorrelated at longer time lags then 
$>~20~$days. This time scale is close to that which is obtained from 
the angular acceleration ($\dot \Omega $) 
 time series. This suggests  
that  
 mass accretion episodes are lasting tens of days 
and  giving a certain angular momentum 
  to the neutron star.  
 Therefore the angular accelerations are  
changing on similar time scales. The way of transfering angular 
momentum gives us information about the accretion proccess.       
 
 The torque 
 on the neutron star ($\tau ~<<~T_{m}$, where $\tau $ and  $T_{m}$ is the
 crust core coupling time and torque 
 measuring time respectively) can be expressed as a specific angular 
 momentum ($l$) added to neutron star at some radius with a certain mass 
 accretion rate (Lamb 1991), 
\begin{equation}
I\dot \Omega = \dot M l. 
\end{equation}  

 If the accretion is from a Keplerian disk   
 (Ghosh $\&$ Lamb 1979a,b) then the external torque is given by 
 \begin{equation}
I\dot \Omega = n(w_{s})  \dot M~l_{K},   
\end{equation}     
where $l_{K} = (GMr_{o})^{1/2}$ is 
 the specific angular momentum added by a Keplerian disk 
 to the neutron star at the inner disk edge   
 $r_{o} \approx 0.5 r_{A}$;   
 $r_{A} \ = (2GM)^{-1/7}
  \mu  ^{4/7} \dot M^{-2/7}  $
 is the Alfven radius; $\mu  $ is the neutron star 
 magnetic moment;   
$n(w_{s}) \approx 1.4 (1-w_{s}/w_{c})/(1-w_{s})$ 
 is a dimensionless 
 function that measures the variation of the accretion torque 
as estimated by the fastness parameter  
$w_{s}
 =\Omega /\Omega _{K}(r_{o}) = 2 \pi P^{-1}  G^{-1/2}  M^{-5/7} 
    \mu ^{6/7} \dot M^{-3/7} $. Here $w_{c}$ is the 
critical fastness parameter at which the accretion 
torque is expected to vanish ($w_{c} \sim 0.35-0.85$  
depending on the electrodynamics of the disk, Lamb 1989). 
 In this model, the torque will cause a spin-up   
if the neutron star is rotating slowly  
 ($w_{s}~<~w_{c}$) in the same 
sense as the circulation in the disk, 
or down, if it is rotating in the opposite sense
 (see Lamb 1991). Even if the 
neutron star is rotating in the same sense as the disk flow, 
the torque will spin-down if it is rotating too rapidly
 ($w_{s}~>>~w_{c}$). In this model one should see positive 
correlation  between angular acceleration ($\dot \Omega $) and 
 mass accretion rate ($\dot M$) if the disk is rotating 
in the same sense as the neutron star. If the 
flow is from Roche Lobe overflow then the accreting material 
carries positive specific angular momentum $l$, 
therefore it is hard to imagine accretion flow 
reversals and hence the spin-up/down torques should be 
correlated with mass accretion rate $\dot M$. 
 Recent numerical 
simulations indicate that capture from winds may lead to strong circulations 
at the magnetospheric boundary (or Alfven surface) 
which reverses its sign quasiperiodically
(Blondin et al., 1990). The formation of 
Keplerian disks at the magnetospheric boundary is possible 
(Lamb 1991). Even if the mass accretion is not changing significantly 
 it is 
possible to see transitions from $+l_{K} $ to $-l_{K}$ 
(or vica versa) and hence 
to observe spin-up/down episodes.       

In general variations in the angular acceleration can be expressed in 
terms of 
variations of mass accretion rate $\delta \dot M$ and
 specific angular momentum $\delta l$   
\begin{equation}
\delta \dot \Omega =\frac{\delta \dot M l + \dot M  \delta l}{I}. 
\end{equation}
 The mutual correlations of angular acceleration, mass accretion rate 
 and specific angular momentum can give information about 
 whether the accretion is due to Roche Lobe overflow or a stellar wind.  
 In Fig 7a,b,c, we represent the angular acceleration, pulse flux 
 (or mass accretion rate) and 
 the ratio of angular acceleration to 
 pulse flux (or specific angular momentum) time series.   

 In Fig. 8 a,b,c, we represent the mutual correlations of
 angular acceleration,  pulse flux, 
 and specific angular momentum.
 As seen from Fig. 8a, there is no clear correlation between 
 X ray flux and angular acceleration. In the case of steady accretion 
 disk  models (Ghosh $\&$ Lamb 1979,b), the higher accretion rate is the 
 higher angular acceleration. But Fig 8a, suggest the opposite case, 
 when the accretion rate is increased angular acceleration is decreased.  
 Furthermore, at all flux values angular acceleration can take 
 both positive and negative sign. In Fig 8b, we represented the 
 specific angular momentum  
  versus pulse flux (or mass accretion). 
 In this case the specific angular momentum goes to zero when the 
 flux increases. For the lower values of flux, specific angular 
 momentum has both positive and negative sign. In the case of 
 Keplerian disks the specific angular momentum depends  
weakly on the X ray flux: 
 $l_{K} \sim  L_{x}  ^{-1/7}$. When the mass accretion rate is increased, 
the Alfven radius gets smaller, therefore 
 a decrease in the specific angular 
momentum is expected. But 
it is unlikely to have zero specific angular momentum 
at higher luminosities. This is suggesting that at higher 
accretion rates the flow geometry is changing. Possibly the accretion 
is becoming radial and dumping the material 
onto the neutron star. On the other hand at lower flux values, 
the specific angular momentum has both positive and negative sign 
and the absolute magnitude of specific angular momentum is increased. 
This suggests that Keplerian disks forms at lower  
accretion rates and these disks can have both positive and negative 
circulations.
 In Fig 8c, we represent the correlation between specific angular 
momentum and angular accelerations. There is a strong correlation between 
specific angular momentum and angular acceleration. 
 The above correlations 
imply that the specific angular momentum is directional, 
 sometimes positive and sometimes negative (or $\pm l$), 
and that sometimes the flow is radial.
These results suggest the formation of
accretion disks in the case of stellar wind accretion  
and the short term disk reversals are quite possible.
 
 In the hydrodynamical simulations of nonaxisymetric gas 
with transverse velocity and density gradients (or accretion 
from inhomogeneous stellar wind) 
 flowing onto a gravitating compact object, the      
formation of a disk  
 and reversals of circulations in the disk are seen 
 (Taam $\&$ Fryxell 1988a,b; 
Taam $\&$ Fryxell 1989; Blondin et al., 1990). These simulations 
also showed that the   
while the disk is present, the specific angular momentum is high 
and the mass accretion rate is low. When
the disk alternates its sense of rotation, the mass accretion increases 
rapidly, leading to flare like activity. At MJD $\sim $ 48773 and 
MJD $\sim $ 49004, the source flux increased rapidly (see Fig. 7a) while the 
source transited from spin-up to spin-down (see Fig. 7b). 
During these flares the specific angular momentum was very low 
(see Fig. 7c) (It should be noted that these flares are the 
 highest flux points in 
Fig. 8a,b.). This is very similar to what is seen in hydrodynamical 
simulations 
(Taam $\&$ Fryxell 1988a,b). In the radiatively driven stellar winds, 
a fraction of the wind is focused by the gravitational 
force of the neutron star and shocked in an accretion 
bow shock. The X-rays emitted from the neutron star heats 
and photoionizes the wind, and create an
 ionized gas at the Str{\"o}mgren zone. 
This effect slows down the velocity of the gravitationally focused 
 material and increases the density of the wind.
This will greatly
enhance
the mass accretion rate onto the neutron star.
The  
slowly moving high density wind can form a Keplerian disk 
closer to the neutron star magnetosphere 
(Blondin et al., 1990).
According to the distribution  
velocity and density gradients, the flow can change its direction 
quasiperiodically.   
The time scale of the flow reversals in the hydrodynamical 
simulations  
are of the order of the free fall time scale ($t_{ff} \sim r_{a}/ v_{rel} $)   
from the capture radius ($r_{a}=2GM/ v_{rel}^{2}$, where 
$v_{rel}$ is the relative velocity of the wind) 
(see also Taam $\&$ Fryxell 1988a,b;
Taam $\&$ Fryxell 1989; Blondin et al 1990).
 This time scale for the slow winds 
can be  $\sim 5~$days (Blondin et al 1990)
 which is close to the spin-down episodes seen  
 in OAO 1657-415 angular velocity time series history. It should also be noted 
that spin-up episodes are longer than the spin-down episodes 
(see Fig. 2). 
This is implying that the slowed down dense wind has some 
specific angular momentum in the sense of orbital motion
 due to the Coriolis force  
and
 this is giving extra spin-up torques rather than spin-down.  


\section{Conclusion}

 In this work, we model the angular velocity 
 time series of OAO 1657-415 and characterize the type of accretion 
 (wind or disk).
 The noise power spectrum and the structure function analysis 
 showed that the angular velocity    
 time series history of OAO 1657-415 is consistent with a random walk 
 model,( with steps at order of several days: $\Delta t \sim 8-19$ days) 
 and that all the observed fluctuations are associated with external 
 torques. The random walk strength   
  $S \sim 8 \times 10^{-16}$ rad$^{2}$/sec$^{3}$ is 
 consistent with that found by 
 Baykal $\& $ \"{O}gelman (1993), 
$(0.4 - 5.9) \times 10^{-15} $ rad$^{2}$/sec$^{3}$, 
using the data given by Nagase (1989).  
 This noise level is a decade higher than 
 Cen X-3 and GX 1+4 (Finger et al., 1994; Chakrabarty 1995)
 and 
 $10^{2},~10^{3},~10^{4}$ times higher than      
  Her X-1, Vela X-1, 
   4U 1626-67 
  (Boynton 1981; Deeter et al., 1989; Chakrabarty 1995). 
 These are the systems in which the noise power density
 has been studied in detail 
  (for the other sources' estimates see Baykal $\&$ \"{O}gelman 1993).
 The magnitude of the angular accelerations in the spin-up/down episodes 
 are consistent with systems spining-up/down 
rapidly such as 
 SMC X-1 (secular spin-up) and GX 1+4 
(secular spin-up before 1980, and secular spin-down after 1984). 
 The high level of angular accelerations in the spin-up/down 
 episodes  
 suggest that the Keplerian disks are present   
 in these episodes. 
 We have found almost no correlation between angular acceleration 
and mass accretion rate. The strong  
correlation between angular acceleration 
 and the specific angular momentum implies that the  
 specific angular momentum is changing its sign randomly, as is seen 
 in hydrodynamical calculations of wind accretion (Blondin et al., 1990).
 Assuming that mass enters the magnetosphere at 
 the corotation radius 
 ($ r_{co} = (GM/ \Omega ^{2} )^{1/3} \sim 
   2 \times 10^{9}~$cm), where the angular momentum 
 is added (or subtracted)  
  according to sign of specific angular momentum 
  ($l_{K} = \pm (GMr_{co})^{1/2}$), and   
  using the mass accretion rate 
$\sim 10^{17}$ gm~sec$^{-1}$ (Chakrabarty et al., 1993), and the approximate 
 values obtained for the torque event rate $(R \sim 18~$days$^{-1})$ 
 and duration of events $(\Delta t \sim 14~$days), we obtained the  
 approximate value of the random walk strength as,  
\begin {equation}
 S=R<(\delta \Omega )^{2} > =
 R <( \dot M \Delta t (G M r_{co})^{1/2}/I )^{2}  > \sim   
 8.6  \times 10^{-16}~{\rm rad^{2}~sec^{-3}}.  
\end{equation} 
This is close to our random walk noise strength value, 
 indicating the possible formation of episodic 
 Keplerian disks. Note however that the mass accretion rate used here is rather uncertain, since
it is based on a luminosity which is calculated indirectly using a value
for the pulsed fraction from earlier observations, and a very uncertain
distance of ~ 10 kpc based mainly on a high interstellar absorption and a low
galactic latitude (Kamata et al. 1990).
Eq. 15 also shows that the level of noise strength increases with 
mass accretion rate.                          

 Our findings are strongly sugesting that OAO 1657-415 has an  
 OB type companion. In the case of a Be type of companion,  
 the accretion disk forms from the  
 low velocity equatorial stellar wind.
 In this case the orbital velocity of the system is higher than  
 the wind velocity.  Therefore 
 the flow has enough initial specific angular momentum, and hence 
 the formed disk should be in the same sign with orbital motion. In this 
 case, it is unlikely to expect disk reversals 
 (or flip flop instabilities) and one should see positive  
 correlations between angular accelerations and mass accretion rates
 (Ghosh $\&$ Lamb 1979a,b).

 One of the important observations to test the idea of 
 formation of disk from the stellar wind is to 
 obtain spectral information and examine the 
 column density (or filaments) during the orbital phases. 
 The distribution of ionization lines and column density during the 
 orbital phases will give additional information 
 about accretion process (Blondin et al., 1990). 
 
\vspace{1cm} 

 \hspace{-1.1cm} {{\bf Acknowledgments}}\\
 It is a pleasure to thank John Deeter, Pranab Ghosh, Jean Swank  
 for stimulating discussions,  
  Mark Finger and   
  CALTECH team, particularly Brian Vaughan,
  Deepto Chakrabarty,  
  Thomas Prince   
  for supplying the pulse frequency and 
  X ray flux records during the work.
 I thank the Compton Gamma Ray Observatory team at HEASARC for 
  the archival data and refree Martijn de Kool for helpful comments. 
 I acknowledge the National 
  Research Council for their support.
 \\

\hspace{-1.1cm}{\Large{\bf References}}

\hspace{-1.1cm} Alpar, M,A., Anderson, P.W., Pines, D., 
                Shaham, J., 1981, ApJ 249, L29

\hspace{-1.1cm} Alpar, M.A., Langer, S.A., Sauls, J.A., 1984, 
                ApJ 282, 533  

\hspace{-1.1cm} Alpar, M.A., Sauls, J.A., 1988, ApJ 327, 723

\hspace{-1.1cm} Alpar, M.A., Chau, H.F., Cheng, K.S., 
                Pines, D., 1993, ApJ 409, 345

\hspace{-1.1cm} Alpar, M.A., 1993  The Lives of Neutron Stars,
                ed, Alpar, M.A., K{\i}z{\i}lo\u{g}lu, \"{U}, van Paradijs, J,
                 NATO/ASI, pg 185

\hspace{-1.1cm} Anzer, U., B\"{o}rner, G., Monaghan, J.J., 
                1987, A$\&$A 176, 
                235

\hspace{-1.1cm} Amstrong, J.T., Johnston, M.D.,Bradt, H.V., Cowley, A.P., 
                Doxsey, R.E., Griffiths, R.E., Hesser, J.E., 
                Schwartz, D.A., 1980, ApJ 236, L131

\hspace{-1.1cm} Ainsworth, T., Pines, D., Wambach, J., 1989, 
                Phys. Lett. B 222, 173

\hspace{-1.1cm} Baykal, A., Alpar, A., K{\i}z{\i}lo\u{g}lu,\"{U}., 1991, 
                A$\&$A 252, 664

\hspace{-1.1cm} Baykal, A., \"{O}gelman, H., 1993,
                A$\&$A 267, 119

\hspace{-1.1cm} Baykal, A., Boynton, P.E., Deeter, J.E., \& Scott, M.,
                1993a, MNRAS 265, 347 

\hspace{-1.1cm} Baykal,A., Anderson, S.F., Margon, B. 1993b,
                AJ 106, 2359

\hspace{-1.1cm} Baykal,A., Swank.,J., 1996, ApJ 460, 470  

\hspace{-1.1cm} Baym, G., Pethick, C., Pines, D., Ruderman, M., 1969, 
                Nat 224, 872

\hspace{-1.1cm} Bryne, P., et al., 1979, IAU Circ., No. 3368

\hspace{-1.1cm} Bryne, P., et al., 1981, ApJ 246, 951 

\hspace{-1.1cm} Blackman, R.B., Tukey,J.W.: 1958, The Measurement of
                Power Spectra, (New York:Dover) 

\hspace{-1.1cm} Blondin, J.M., Kalmann, T.R., Fryxell, B,A., 
                Taam, R.E., 1990, ApJ 356, 591

\hspace{-1.1cm} Boynton, P.E.: 1981, in: Pulsars, IAU Symposium No. 95, eds
                  W. Sieber and R. Wielebinski, (Dordrecht: Reidel),  279

\hspace{-1.1cm} Boynton, P.E., Deeter, J.E., Lamb, F.K., Zylstra, G., 
                Pravdo, S.H., White, N.E., Wood, K.S., Yentis, D.J., 
                1984, ApJ 283, L53

\hspace{-1.1cm} Chakrabarty, D.,
                 et al., 1993,  ApJ 403, L33

\hspace{-1.1cm} Chakrabarty, D.: 1995, PhD thesis,
                 California Institute of Technology 

\hspace{-1.1cm} Chau, H.F., McCulloch, P.M., Nandkumar, R., Pines, D., 1993,
                ApJ, 413 L113 

\hspace{-1.1cm} Corbet, R.D.H., 1986, MNRAS 220, 1047 

\hspace{-1.1cm} Cordes, J.M., 1980, ApJ 237, 216

\hspace{-1.1cm} Cordes, J.M., Downs, G.S., 1985, ApJS  
                 59, 343

\hspace{-1.1cm} Datta, B., Alpar, M.A., 1993, A$\&$A  
                275, 210 

\hspace{-1.1cm}  de Kool, M., Anzer, U., 1993, MNRAS, 
                 262, 726  

\hspace{-1.1cm} Deeter, J.E. 1981, PhD thesis, University of Washington 

\hspace{-1.1cm} Deeter, J.E., 1984, ApJ 281, 482

\hspace{-1.1cm} Deeter, J.E., Boynton, P.E., 1982, ApJ 261, 337

\hspace{-1.1cm} Deeter,J.E., Boynton,P.E.,
                Lamb,F.K., Zylstra,G. 1989., ApJ
                336, 376

\hspace{-1.1cm} Easson, I., 1979, ApJ 228, 257 

\hspace{-1.1cm} Finger, M.H, Wilson, R.B., $\&$ Fishman, G.J., 1994,
                In Second Compton Symposium, ed. C.E.Fichtel, N.Gehrels, 
                $\&$ J.P.Norris (New York: AIP Press), 304 

\hspace{-1.1cm} Finger, M., Wilson, R.B., $\&$ Harmon, B.A., 1996, 
                ApJ in press 

\hspace{-1.1cm} Fishman, G.J., et al., 1989, in Proc. GRO 
                Science Workshop, ed. W.N. Johnson (Greenbelt: NASA/GSFC)
                pg 2 

\hspace{-1.1cm} Ghosh, P., Lamb, F.K., 1979a, ApJ 232, 259

\hspace{-1.1cm} Ghosh, P., Lamb, F.K., 1979b, ApJ 234, 296

\hspace{-1.1cm} Gilfanov, M., Sunyaev, R., Churazov, E., Babalyan, G., 
                Pavlinskii, M., Yamburenko, N., Khavenson, N., 1991, 
                Soviet Astron. Lett., 17, 46

\hspace{-1.1cm} Jenkins, G.M., Watts, D.G., 1968, Spectral Analysis 
                And Its Applications, Holden Day  

\hspace{-1.1cm} Kamata, Y., Koyama, K., Tawara, Y., Makishima, K., 
                Ohashi, T., Kawai, N., Hatsukade, I., 1990, PASJ 
                42, 785

\hspace{-1.1cm} Lamb, F.K., Pines, D., Shaham, J., 1978a, ApJ 
                224, 969 

\hspace{-1.1cm} Lamb, F.K., Pines, D., Shaham, J., 1978b, ApJ
                225, 582

\hspace{-1.1cm} Lamb, F.K., 1991,
                Frontiers of Stellar Evolution, ed by D.L.Lambert
                (Astronomical Society of the Pacific), pp 299-388 (1991)

\hspace{-1.1cm} Lamb, F.K., 1989, Timing Neutron Stars, ed. 
                H. \"{O}gelman and E.P.J. van den Heuvel (Dordrecht: 
                Kluwer), p. 649.   

\hspace{-1.1cm} Lindsey, W.C., Chie, C.H., 1976, IEEE 64, 1652

\hspace{-1.1cm} Lyne, A.G., 1993, The Lives of Neutron Stars,  
                ed, Alpar, M.A., K{\i}z{\i}lo\u{g}lu, \"{U}, van Paradijs, J, 
                 NATO/ASI, pg 167

\hspace{-1.1cm} Matsuda, T., Inoue, M., Sawada, K., 1987, 
                MNRAS 226, 785    

\hspace{-1.1cm} Mereghetti, S., et al., 1991, ApJ 366, L23

\hspace{-1.1cm} Nagase, F., 1989, PASJ 41, 1

\hspace{-1.1cm} Parmar, A., et al., 1980, MNRAS 193, 49P

\hspace{-1.1cm} Parmar, A., et al. 1989, ApJ 338, 359

\hspace{-1.1cm} Polidan, R.S., Pollard, G.S.G., Sanford, P.W., Locke, M.C., 
                1978, Nat 275, 296

\hspace{-1.1cm} Prince, T.A., Bildsten, L., Chakrabarty, D., 
                Wilson, R.B., Finger, M.H., 1994, In Evolution of 
                X-Ray Binaries, ed. S.S.Holt and C.S.Day (New York:AIP 
                Press), 235

\hspace{-1.1cm} Rice, S.O., 1954, Selected Papers on Noise and 
                Stochastic Processes. Wax N. (ed.), Dover, 
                London, p. 133 

\hspace{-1.1cm} Rutman, J., 1978, Proc. IEEE 66, 1048

\hspace{-1.1cm} Sauls, J.A., 1988 Timing Neutron Stars, ed, 
                \"{O}gelman,H., van den Heuvel, E.P.J,  NATO/ASI,     

\hspace{-1.1cm} Scargle, J.D., 1981, ApJS 45, 1

\hspace{-1.1cm} Shemar, S.L., The Lives of Neutron Stars,
                ed, Alpar, M.A., K{\i}z{\i}lo\u{g}lu, \"{U}, van Paradijs, J,
                 NATO/ASI, pg 177

\hspace{-1.1cm} Shapiro, S.L., Lightman, A.P., 1976, ApJ 
                 204, 555 

\hspace{-1.1cm} Sunyaev, R., Gilfanov, M., Goldurm, A.,
                Schmitz-Frayesse, M.C.,
                1991, IAU Circ., No. 5342

\hspace{-1.1cm} Taam, R.E., Fryxell, B.A., 1988a, ApJ 
                327, L73 

\hspace{-1.1cm} Taam, R.E., Fryxell, B.A., 1988b, ApJ 
                 335, 862 

\hspace{-1.1cm} Taam, R.E., Fryxell, B.A., 1989, ApJ 
                339, 297 

\hspace{-1.1cm} Waters, L.B.F., van Kerkwijk, M.H., 1989, 
                A$\&$A 223, 196 

\hspace{-1.1cm} White, N.E., Pravdo, S.H., 1979, ApJ 233, L121

\hspace{-1.1cm} White, N.E.., Swank, J.H., Holt, S.S., 1983, ApJ 
                270, 711 

\hspace{-1.1cm} Wilson, R.B., Finger, M.H., Pendleton, G.N., 
                Briggs, M., Bildsten, L., 
                1994, In Evolution of
                X-Ray Binaries, ed. S.S.Holt and C.S.Day (New York:AIP
                Press), 475 


\vspace{1cm}

\hspace{-1.1cm} {{\bf Figure Caption }}\\

{\bf Fig.1} Angular velocity time series history
                   of OAO 1657-415. \\ 

{\bf Fig.2} Angular velocity records of BATSE observations
              (same as Fig.1). \\

{\bf Fig.3} Power spectrum of angular accelerations
                (or noise strenghts, see the text).
            The asterisks denote the measurement errors. \\

{\bf Fig.4} Autocorrelation function of angular accelerations 
            time series.\\

{\bf Fig.5}  a) Theoretical angular accelerations for a crust
         core coupling time $\tau =1~$day and a 
         ratio of core superfluid 
          moment of inertia to crust 
         moment of inertia $I_{s}/I_{c}= 0,3,10$.
         b) Simulated angular accelerations
         (with larger error bars) with $\tau =1~$day and
         $I_{s}/I_{c}=10$ and observed angular accelerations.
         c) Simulated angular accelerations for pure random walk
         model (larger error bars) and observed angular accelerations
         (see also text).\\

{\bf Fig.6 } Autocorrelation function of pulse flux time series.\\

{\bf Fig.7 } BATSE observations of {\bf a)} pulse flux, 
            {\bf b)} angular 
             acceleration (horizontal line denotes 
             the secular spin-up rate), {\bf c)}
             angular acceleration/flux
             (or specific angular momentum) time series
              (Note that angular accelerations are obtained in 
              a time span, as an inverse of the crossover frequency 
              $\sim 8~$days ). \\

{\bf Fig.8 } Correlations between {\bf a)} angular acceleration and 
              pulse flux, 
              {\bf b)} angular acceleration/flux (or specific angular 
               momentum) and pulse flux, {\bf c)} angular acceleration/flux 
               (or specific angular momentum) 
              and angular acceleration (see also text).\\   

   \begin{figure}
\plotone{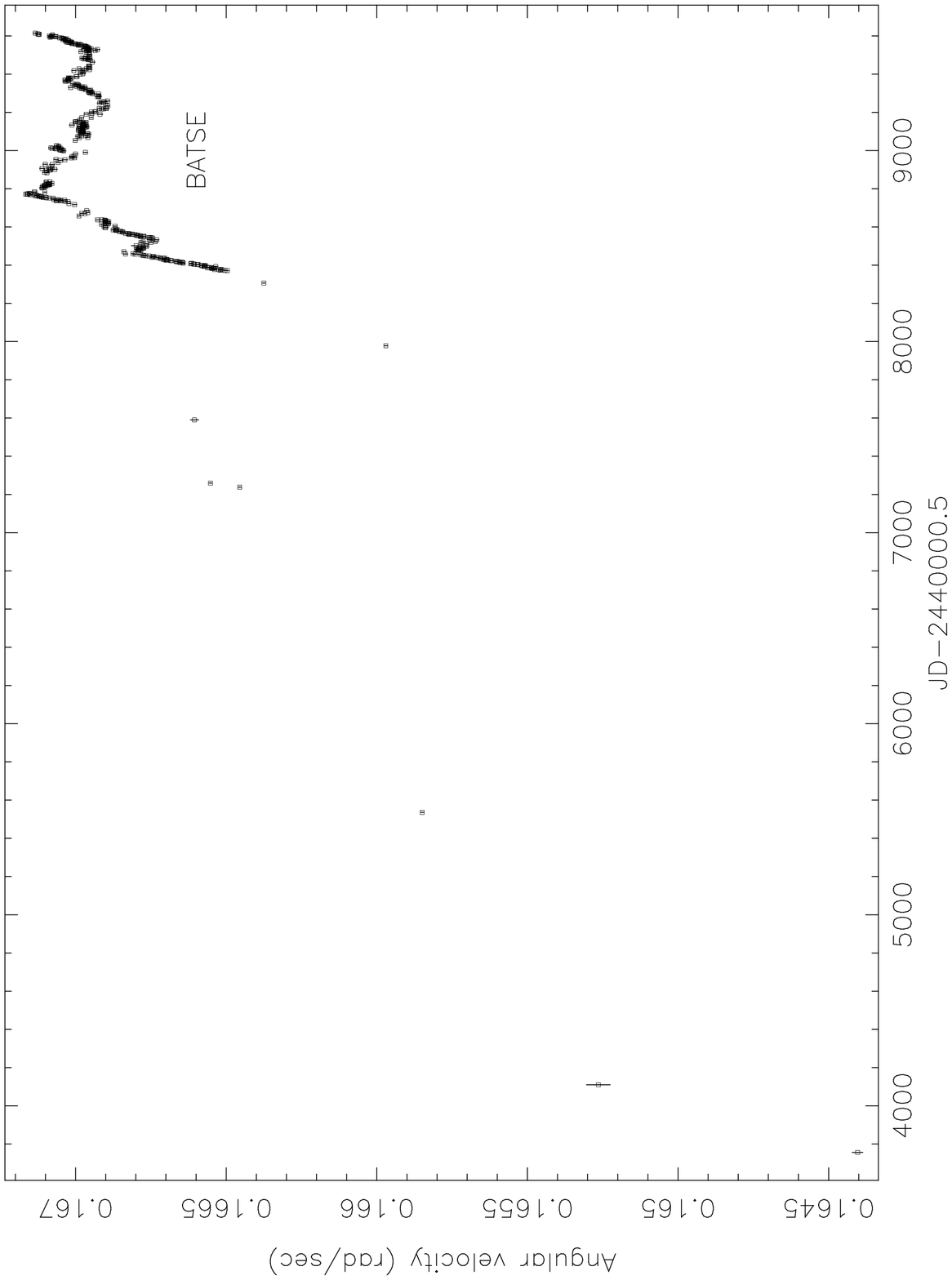}
      \caption{ Angular velocity time series history 
                   of OAO 1657-415   
              }
    \end{figure}
%
%
   \begin{figure}
\plotone{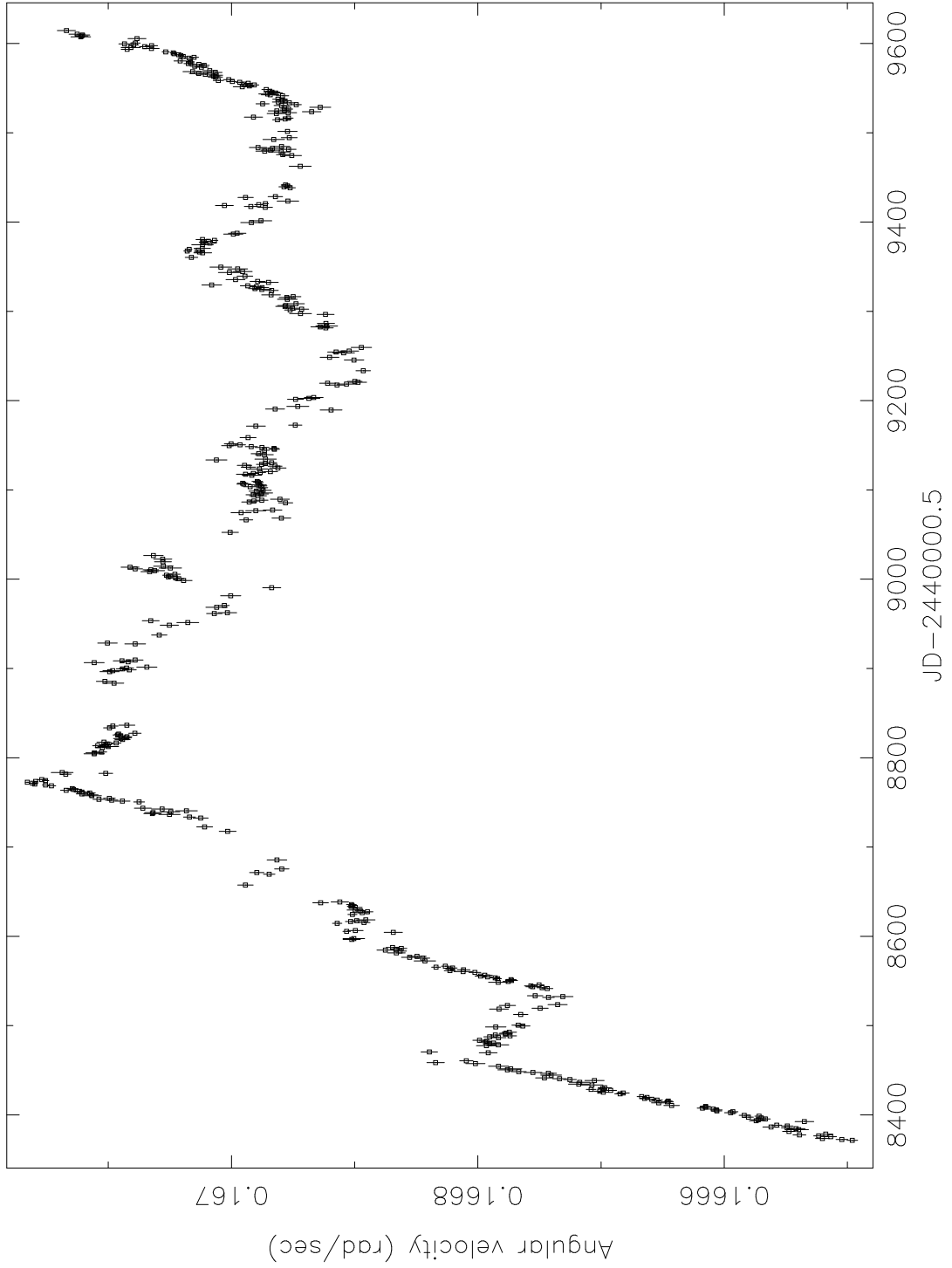 
}
      \caption{ Angular velocity records of BATSE observations 
              (same as Fig.1) 
              }
    \end{figure}
%
%
   \begin{figure}
\plotone{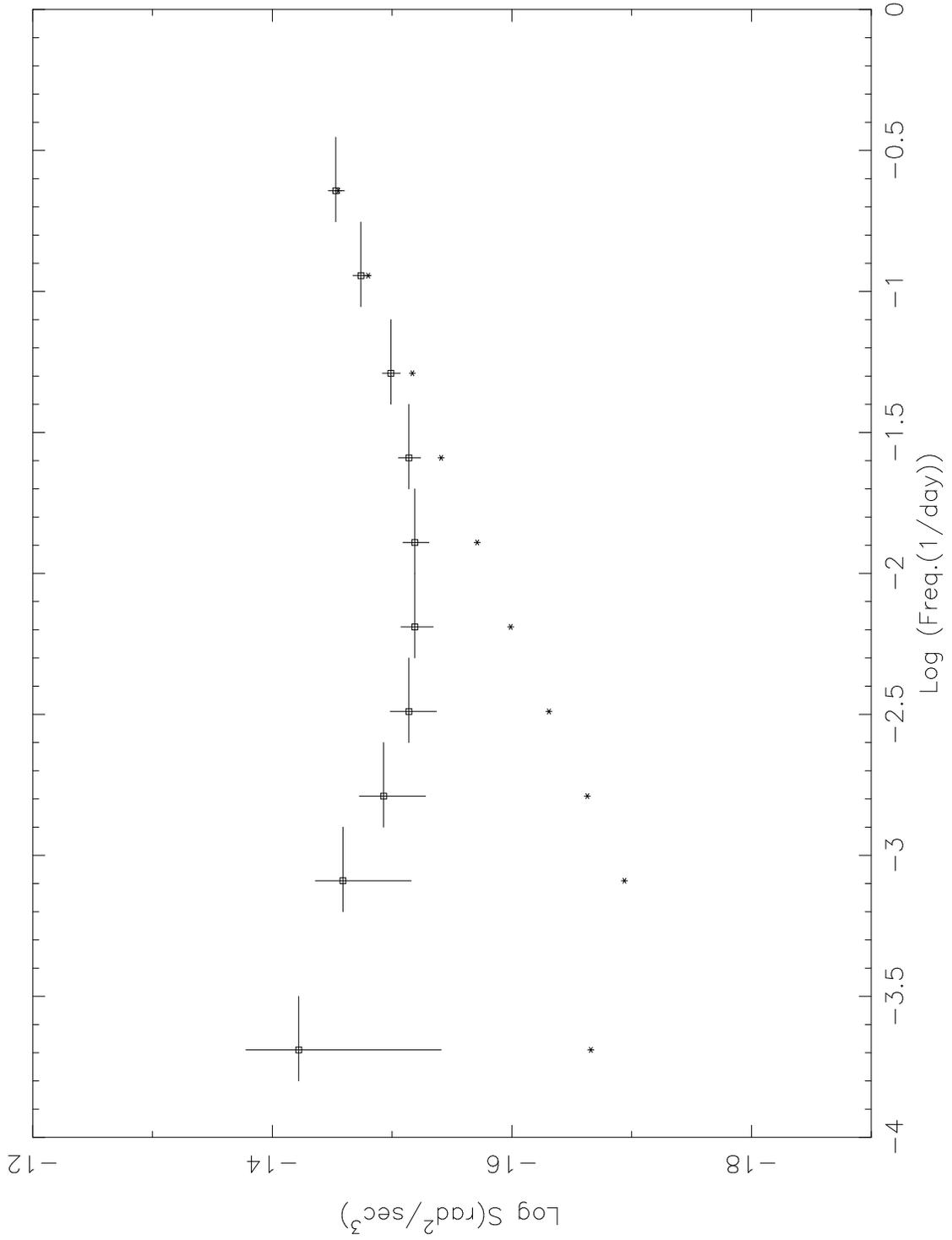
}
      \caption{ Power Spectrum of angular accelerations 
                (or noise strenghts, see the text) 
              }
    \end{figure}
%
%
   \begin{figure}
\plotone{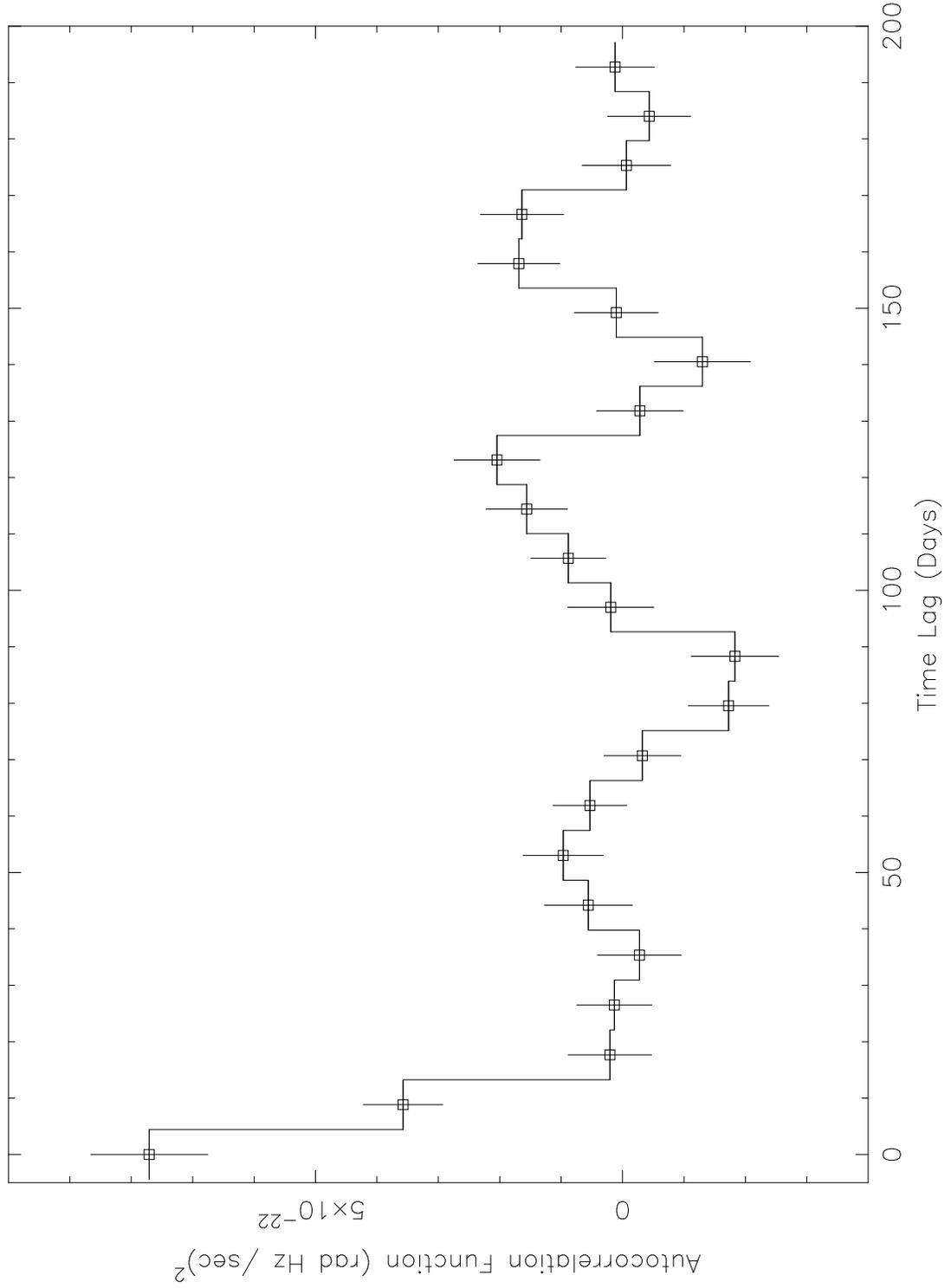
}
      \caption{Autocorrelation function of angular accelerations 
           time series.    }
    \end{figure}
%
%
   \begin{figure}
\plotone{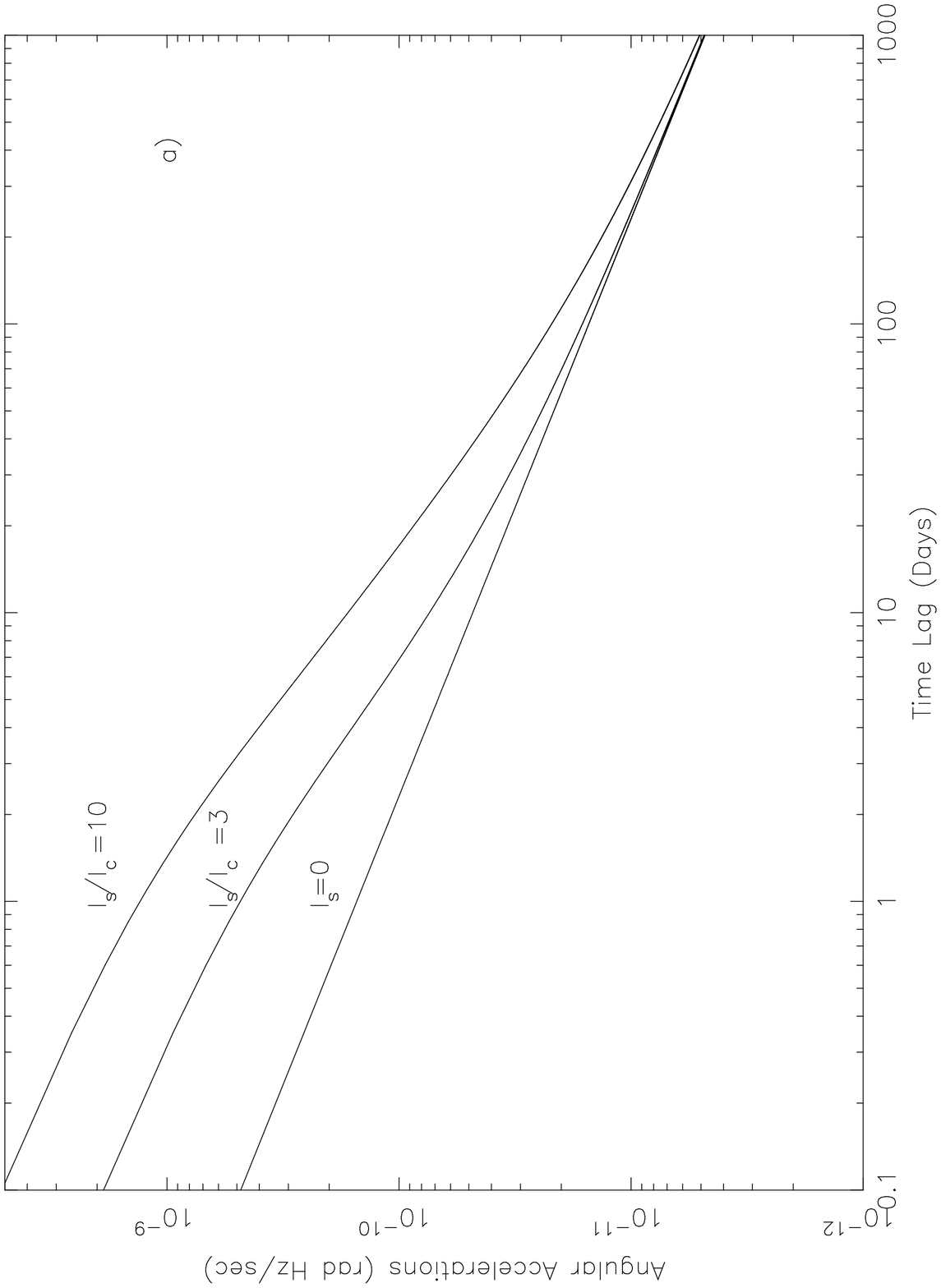
}
    \end{figure}
%
%
   \begin{figure}
\plotone{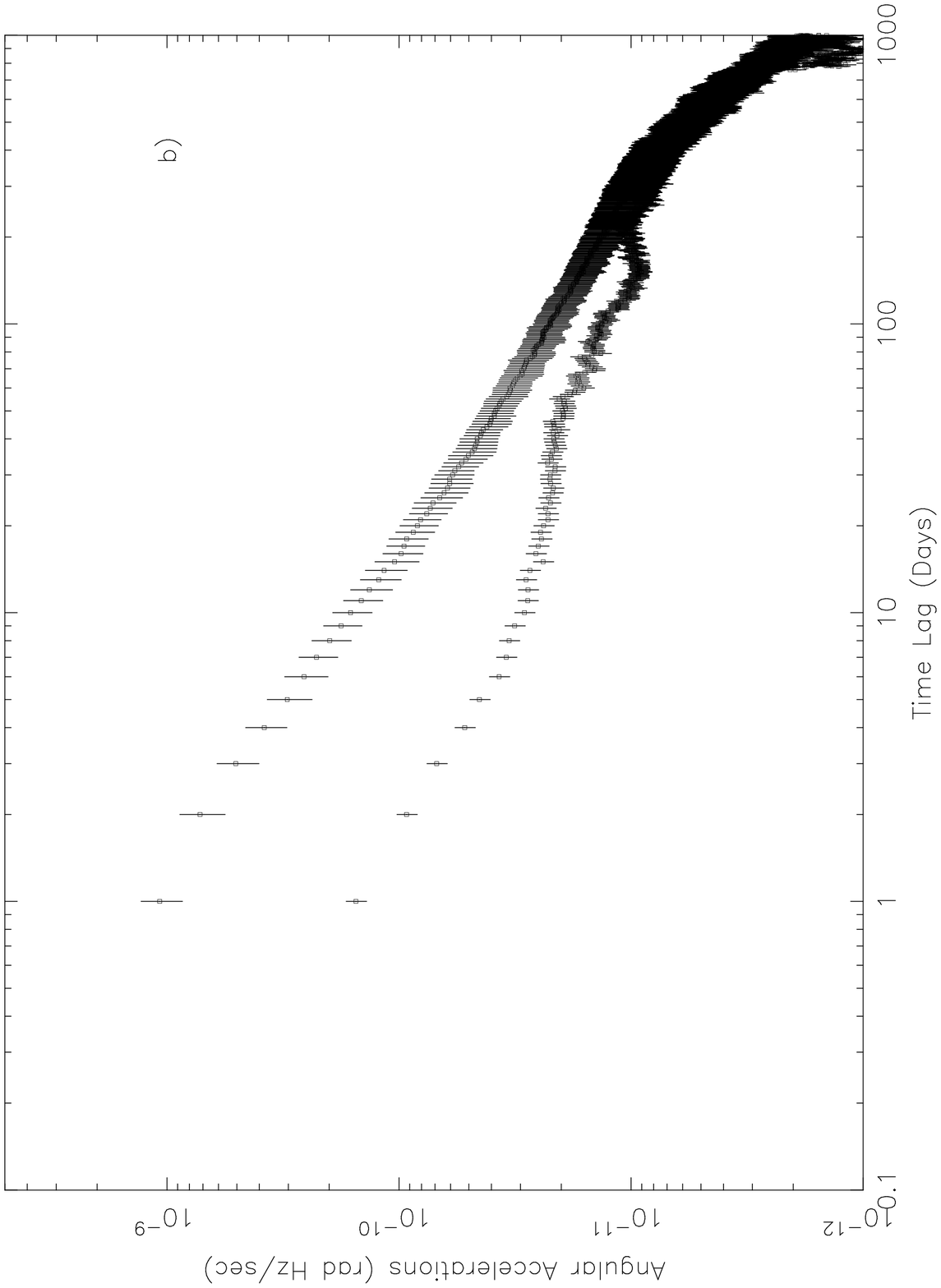
}
    \end{figure}
%
%
   \begin{figure}
\plotone{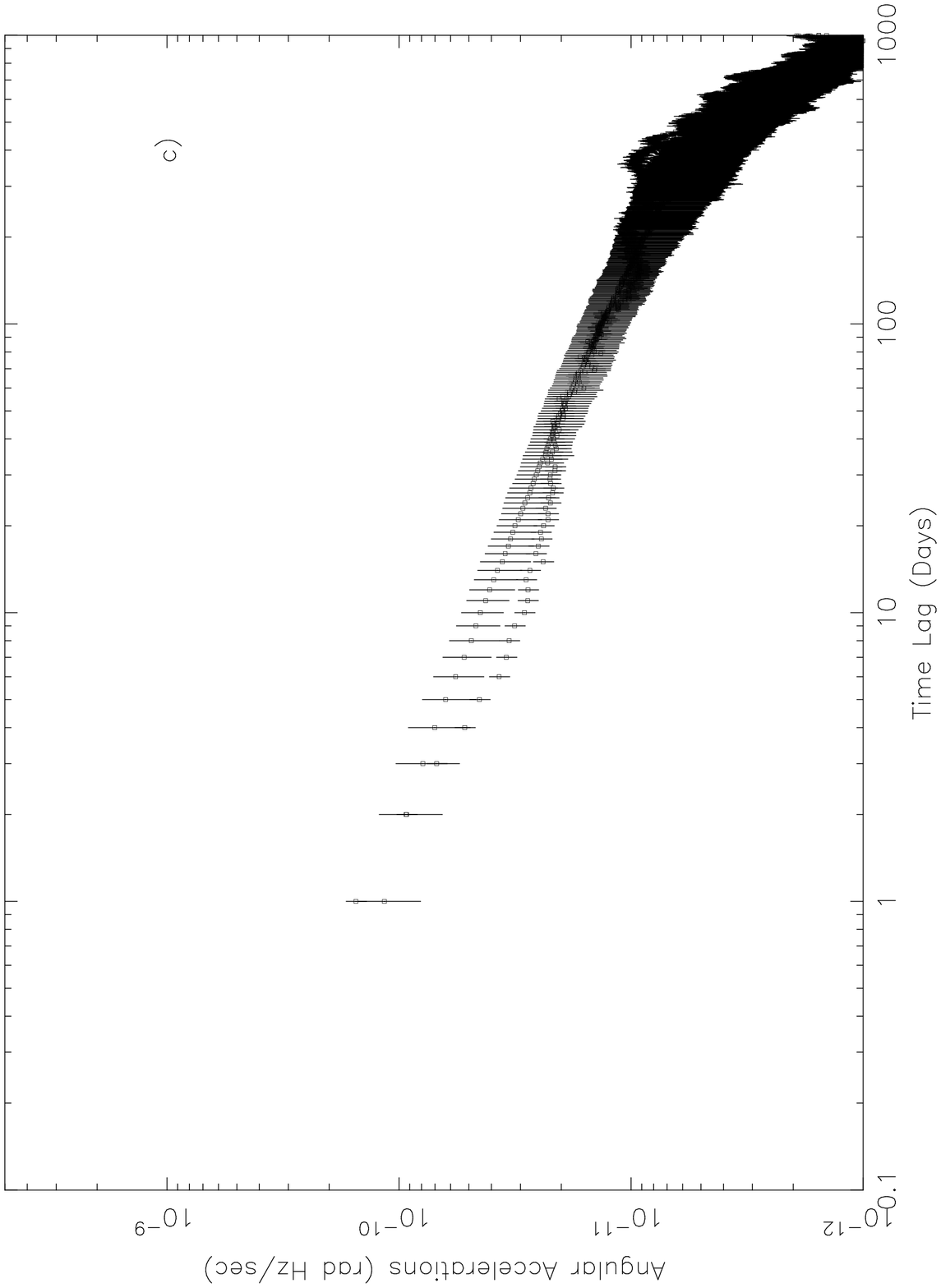
}
      \caption{
              }
    \end{figure}
%
%
   \begin{figure}
\plotone{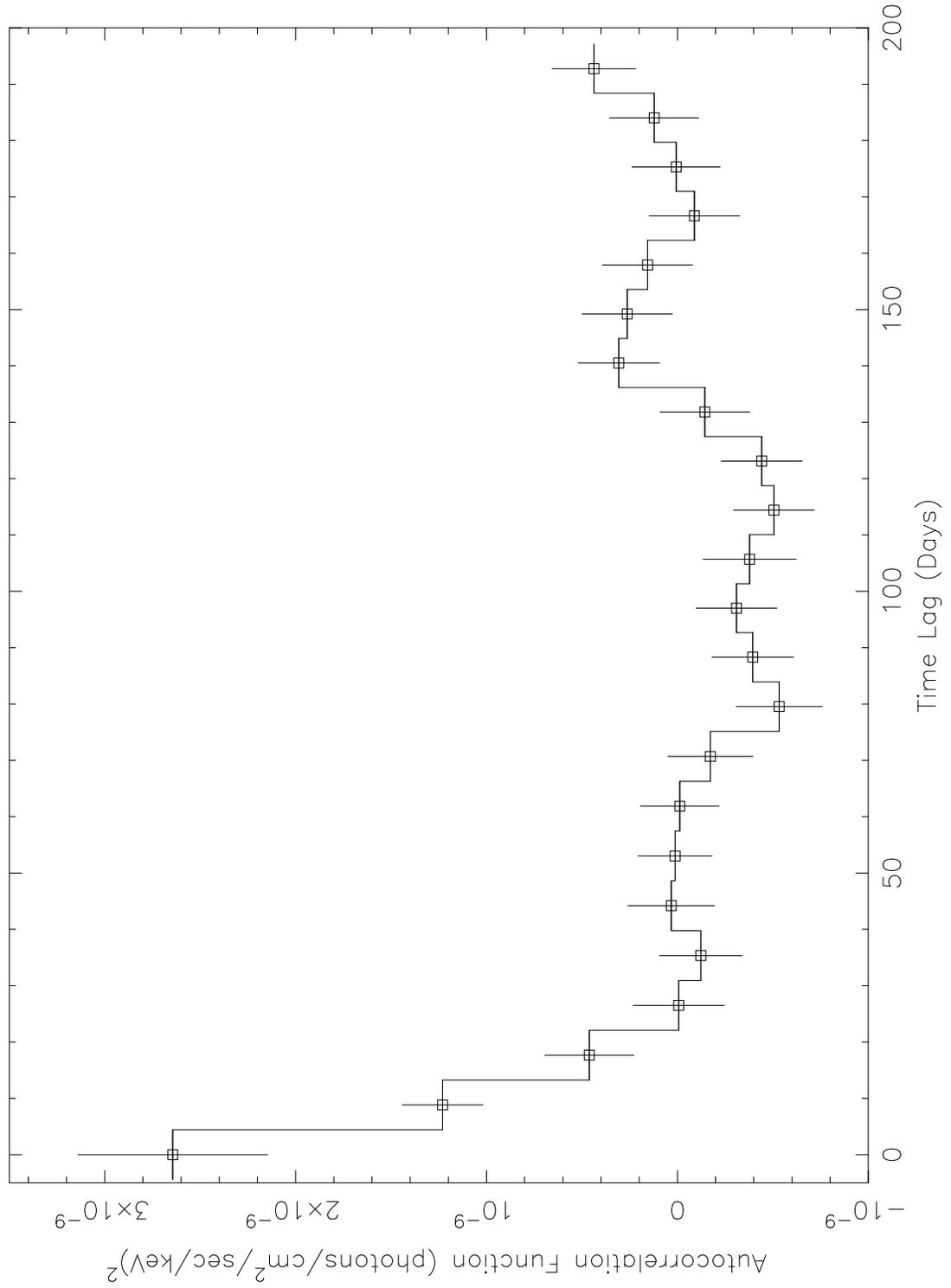
}
      \caption{Autocorrelation function of pulse flux time series 
              }
    \end{figure}
%
%
   \begin{figure}
\plotone{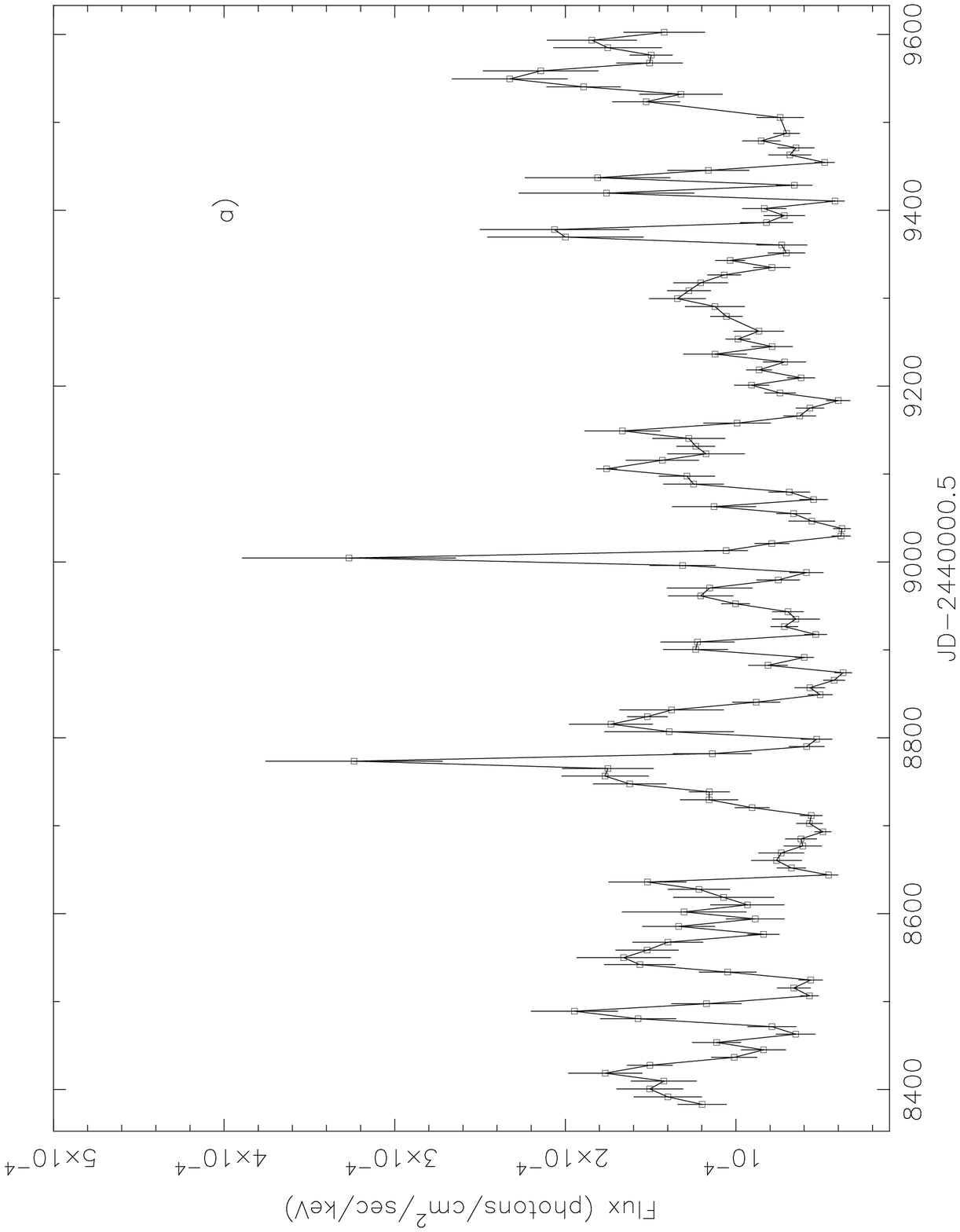
}
    \end{figure}
%
%
   \begin{figure}
\plotone{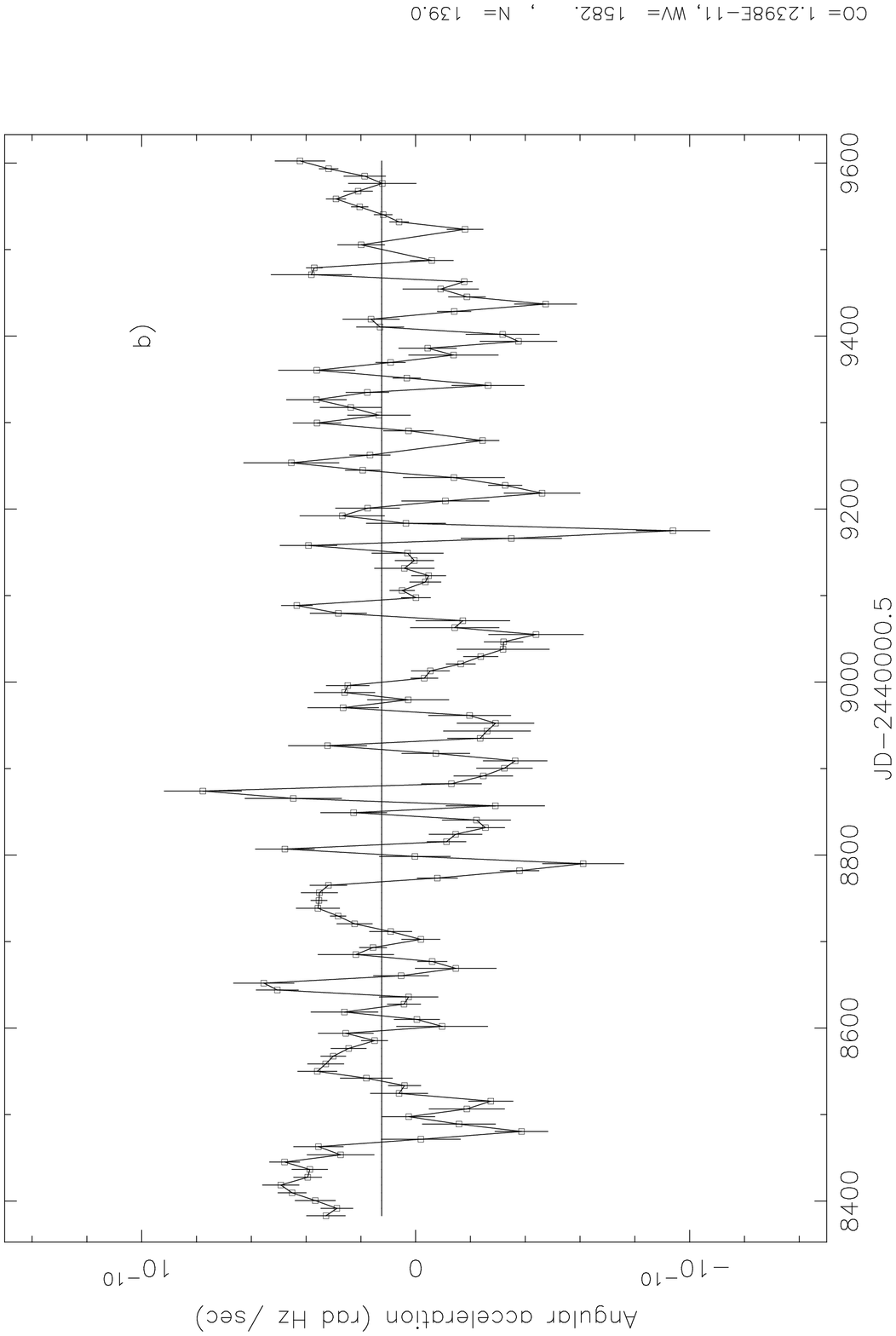
}
    \end{figure}
%
%
   \begin{figure}
\plotone{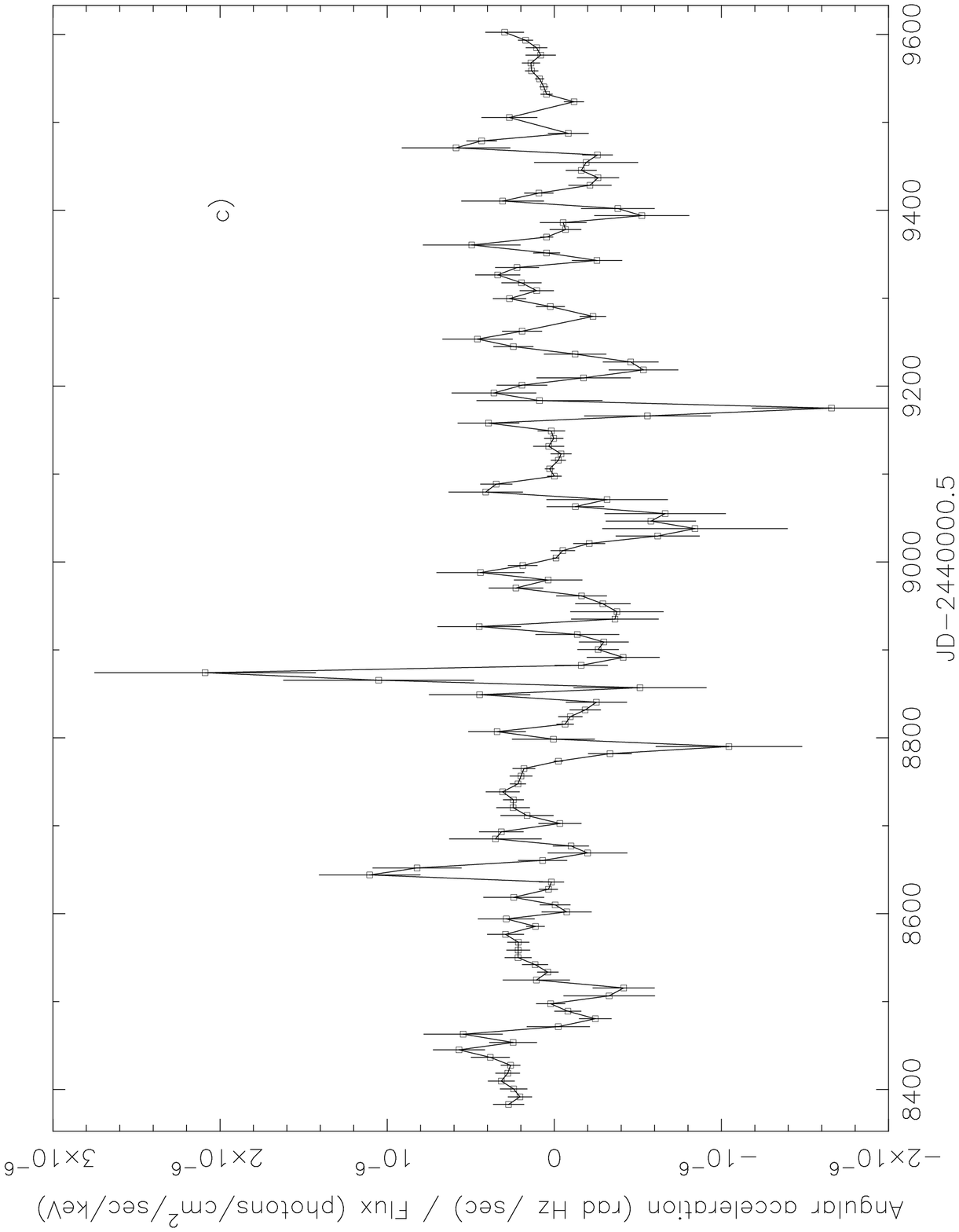
}
      \caption{
              }
    \end{figure}
%
%
   \begin{figure}
\plotone{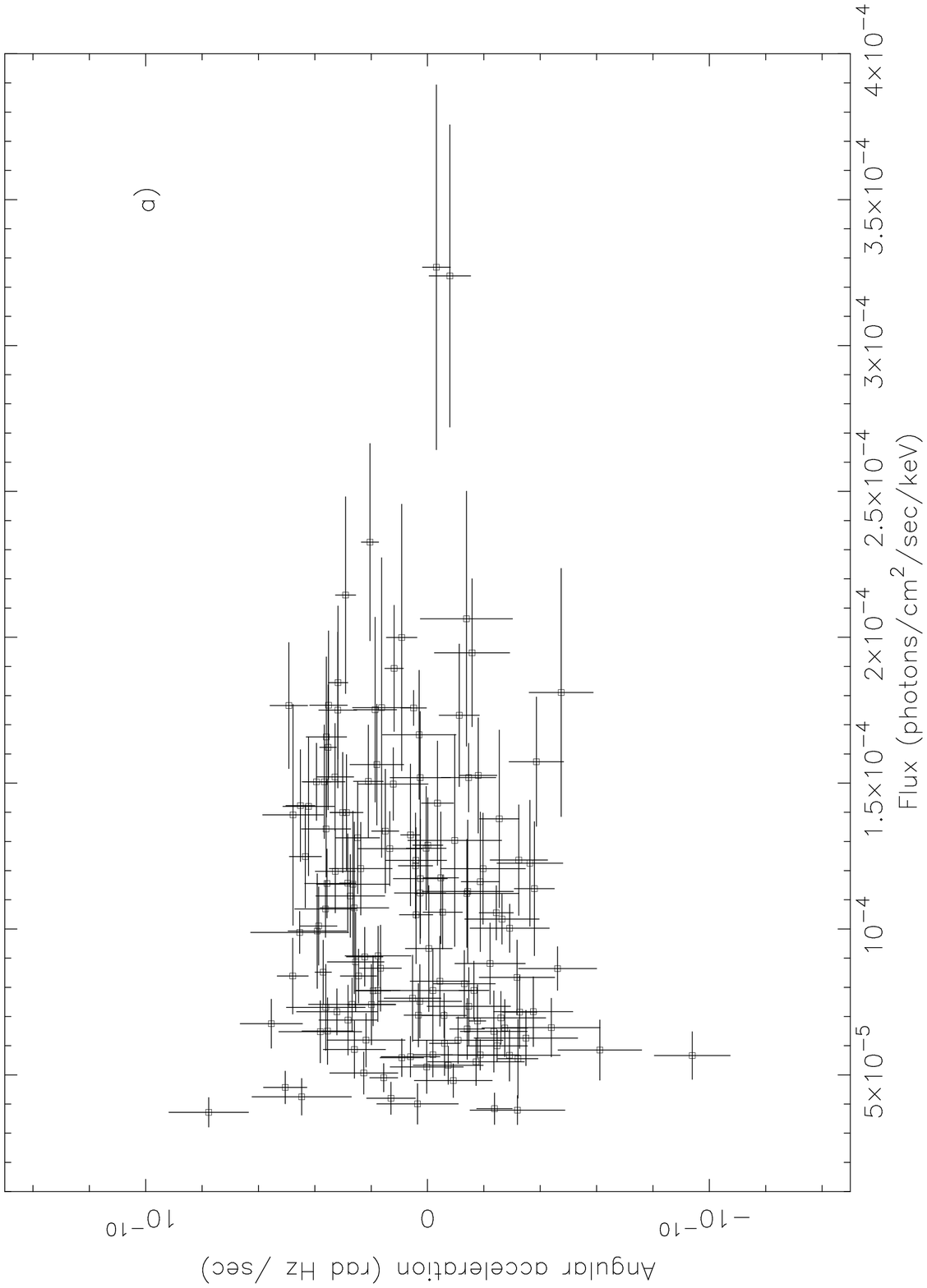
}
    \end{figure}
%
%
   \begin{figure}
\plotone{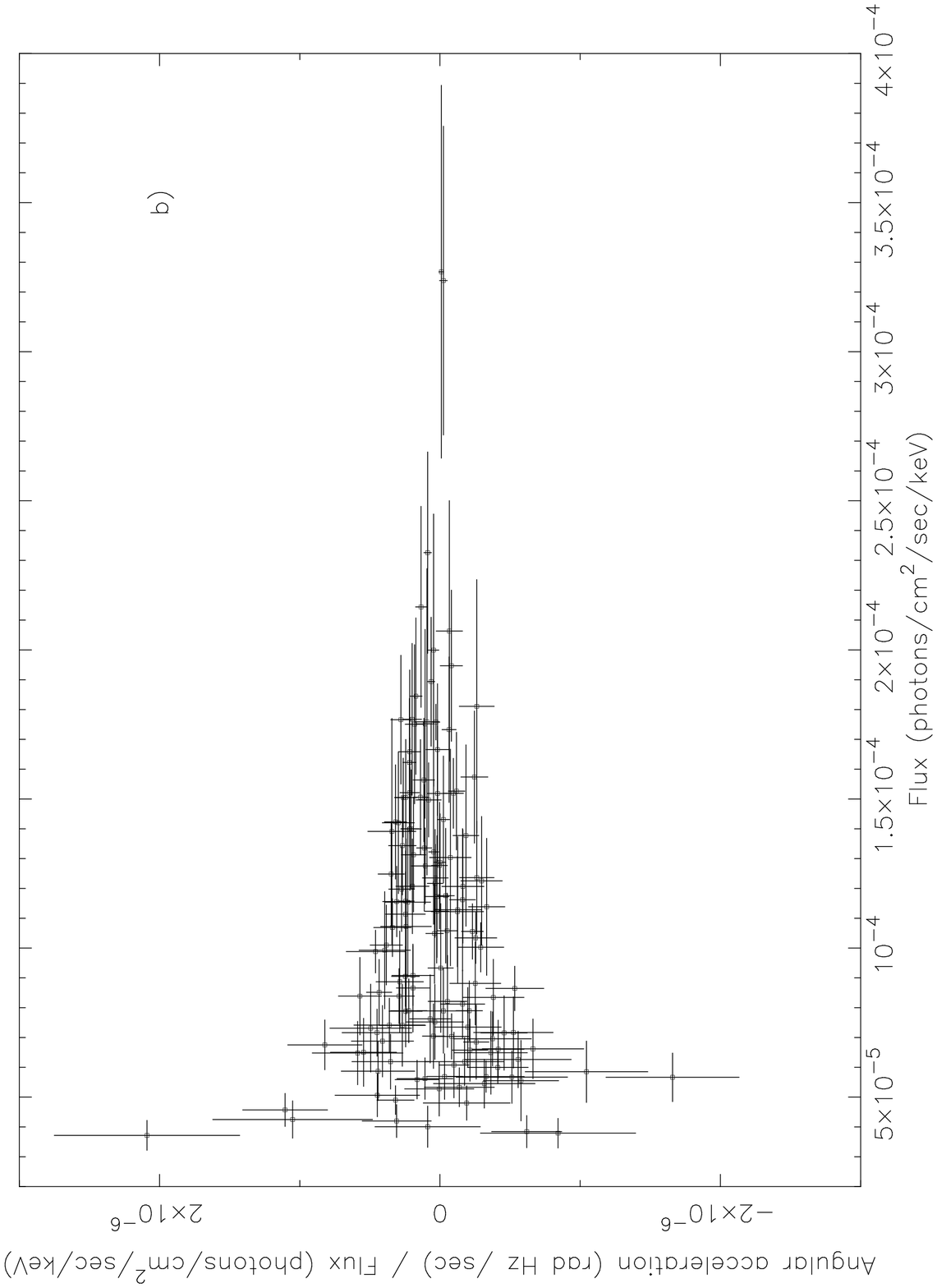
}
    \end{figure}
%
%
   \begin{figure}
\plotone{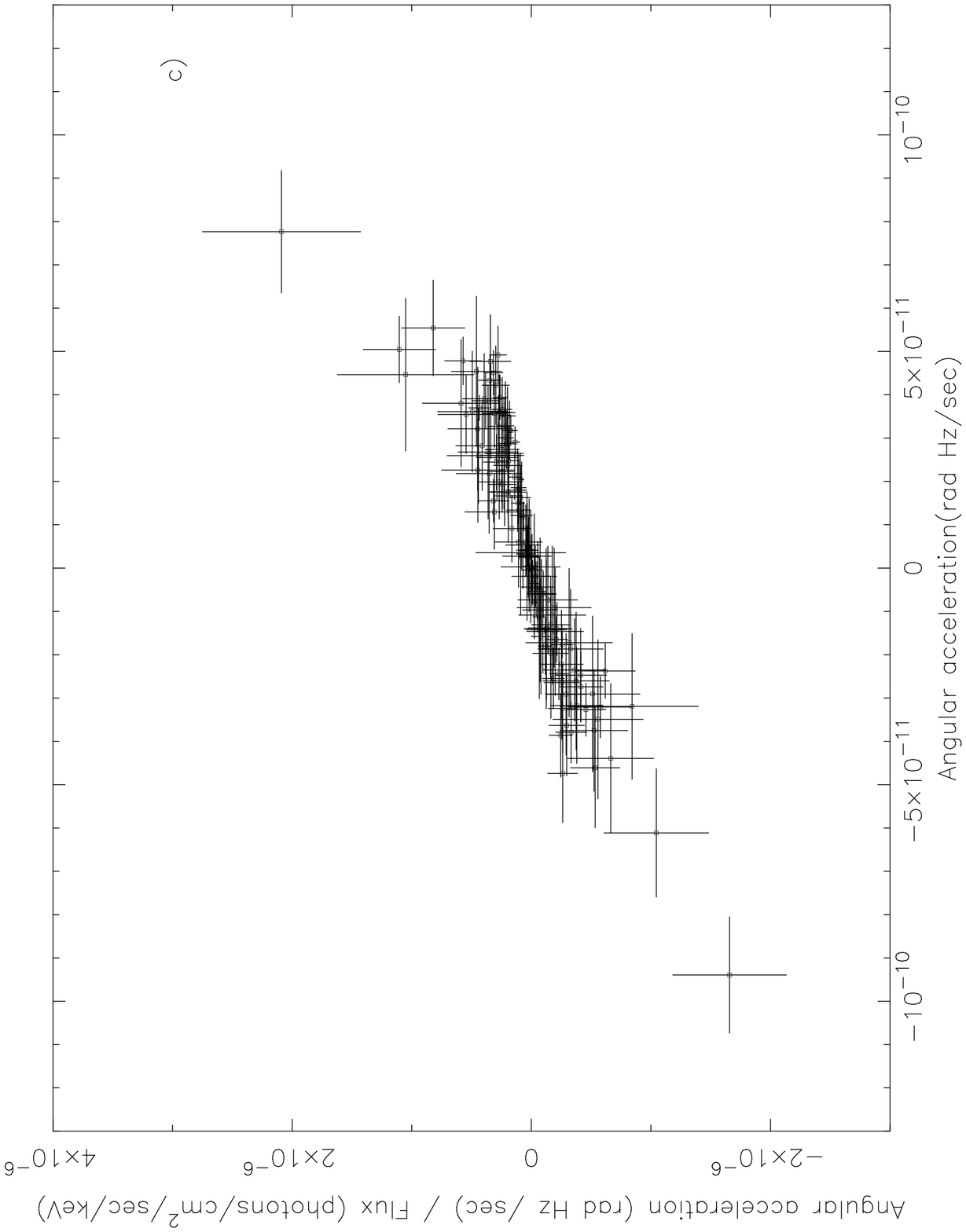
}
      \caption{
              }
    \end{figure}
%
%

\end{document}